\documentclass[twocolumn,showpacs,preprintnumbers,amsmath,amssymb]{revtex4}
\usepackage{graphicx}
\usepackage{dcolumn}% Align table columns on decimal point
\usepackage{bm}% bold math

\begin{document}

\title{
Device Model for   Graphene Bilayer
Field-Effect Transistor
}
\author{V.~Ryzhii\footnote{Electronic mail: v-ryzhii(at)u-aizu.ac.jp}, 
M.~Ryzhii, and A.~Satou
}
\address{
Computational Nanoelectronics Laboratory, University of Aizu, 
Aizu-Wakamatsu, 965-8580, Japan\\
Japan Science and Technology Agency, CREST, Tokyo 107-0075, Japan
}
\author{T.~Otsuji}
\address{
Research Institute for Electrical Communication,
Tohoku University,  Sendai,  980-8577, Japan\\
Japan Science and Technology Agency, CREST, Tokyo 107-0075, Japan
}
\author{N.~Kirova}
\address{Laboratoire de Physique des Solides, Univ. Paris-Sud, 
CNRS, UMR 8502, 
F-91405 Orsay Cedex, France.
}
\date{\today}
\begin{abstract}
We present an analytical device model for a   graphene bilayer field-effect
transistor (GBL-FET) with a graphene bilayer  as a 
channel, and with back and top gates.
The model accounts for the dependences of the electron and hole 
Fermi energies
as well as  energy gap in different sections of the channel
on the bias back-gate and top-gate voltages.
Using this model,
we  
calculate the  dc and ac source-drain currents
and the transconductance of GBL-FETs
with both ballistic and collision dominated electron transport
as functions of structural parameters, 
the bias back-gate and top-gate voltages, and the signal frequency.
It is shown that  there are two threshold voltages, 
$V_{th,1}$ and $V_{th,2}$,
 so that
the dc current versus  the top-gate voltage relation
markedly changes depending on whether the section of the channel
beneath the top gate (gated section) is filled with electrons,
depleted, or filled with holes.
The electron scattering leads to  a decrease in the dc and ac 
currents and  transconductances, 
whereas it weakly affects the threshold frequency.
As demonstrated, the transient recharging of the gated section by holes 
can pronouncedly influence the ac transconductance
resulting in its nonmonotonic  frequency dependence with a maximum
at fairly high frequencies.
\end{abstract}
\maketitle
%\newpage
\section{Introduction}

The features of the electron and hole 
energy spectra 
in graphene provide the
exceptional properties of
graphene-based heterostructures and devices~\cite{1,2,3,4,5,6}.

However, due to the gapless energy spectrum,
the interband tunneling ~\cite{7} can substantially deteriorate
the performance of graphene field-effect transistors (G-FETs)
with realistic device structures~\cite{8,9,10,11}.
To avoid drawbacks of the characteristics of 
G-FETs
based on graphene monolayer with zero energy gap,
the patterned graphene (with an array of graphene nanoribbons)
and the graphene  bilayers can be used in
graphene nanoribbon FETs (GNR-FETs) and in graphene bilayer FETs (GBL-FETs),
respectively.
The source-drain current in GNR-FETs and GBL-FETs, as in the standard FETs,
depends on
the gate voltages. 
The positively biased  back gate provides the formation
of the electron channels, whereas the negative bias
voltage  between the top gate and the channels results in
forming a potential barrier for electrons which controls the 
current.
By properly choosing the width of the nanoribbons, one can fabricate graphene structures with a relatively wide band gap~\cite{12}
(see also Refs.~\cite{13,14,15,16})
Recently, the device dc and ac characteristics of
GNR-FETs were assessed using both numerical~\cite{14} 
and analytical~\cite{17,18,19} models.
The effect of the transverse electric field (to the GBL plane) 
on the energy spectrum
of GBLs~\cite{20,21,22} can  also be used to manipulate and optimize
the GBL-FET characteristics.
A significant feature of GBL-FETs is that 
under the effect of the transverse 
electric field not only the density of the two-dimensional 
electron gas  in the GBL varies, but the energy  gap between 
the GBL valence and conduction bands appears.
This effect can markedly influence the GBL-FET characteristics.
The structure of a GBL-FET is
shown in Fig.~1.
In this paper, we present a simple analytical device model for  a
GBL-FET,  
obtain the device dc and ac characteristics, and compare 
these characteristics
with those of GNR-FETs. 

The paper organized as follows.
In Sec.~II, we consider the GBL-FET band diagrams
at different bias voltages and estimate the energy gaps
and the Fermi energy 
in different sections of the device. Section~III deals with
the Boltzmann kinetic equation which governs the electron
transport at dc and ac voltages and the solutions
of this equation. The cases of the ballistic
and collision dominated electron transport are considered.
In Sec.~IV and Sec.~V, the dc transconductance and the ac frequency-dependent
transconductance  are calculated
using the results of Sec.~III. 
Section~VI deals with the  demonstration  and analysis of
the main obtained results, numerical estimates, and comparison of the GBL-FET
properties with those of GNR-FETs.
In Sec.~VII, we draw the main conclusions.
In Appendix, some intermediate calculations related to
the dynamic recharging of the gated section by holes due 
to the interband tunneling are singled out.

\section{GBL-FET Energy band diagrams}

We assume that the bias  back-gate voltage $V_b > 0$, while the  bias 
top-gate voltage $V_t < 0$. 
The electric potential of the channel at the source and drain
contacts are
$\varphi = 0$ and $\varphi = V_d$, respectively, 
where $V_d $ is the bias drain voltage.
The former results in the formation of a 2DEG in the GBL.
The distribution of the electron density $\Sigma$  along  the GBL 
is generally nonuniform due to the negatively biased top gate
forming the barrier region beneath this gate.
Simultaneously, the energy gap $E_g$ is also a function
of the coordinate $x$ (its axis is directed in the GBL plane from
the source contact to the drain contact) being different 
in the source, top-gate, and drain sections of the channel (see Fig.~2).
Since the net top-gate voltage apart from the bias component $V_t$
comprises the ac signal component $\delta V(t)$,
the height of the barrier for electrons entering 
the section of the channel under the top gate (gated section) 
from the source side
can be presented as

\begin{equation}\label{eq1}
\Delta(t) = \Delta_0  + \delta\,\Delta(t).
\end{equation}

Depending on the Fermi energy  in the extreme sections of the channel,
in particular, on its value, $\varepsilon_F$,
 in the source section and on the height of the barrier 
in this section $\Delta_0$, there are three situations.
The pertinent 
the GBL-FET energy band diagrams are demonstrated in Fig.~2.
The spatial distributions of electrons and holes in the GBL channel
are different depending on the relationship between the top-gate voltage $V_t$
and two threshold voltages, $ V_{th,1}$ and  $ V_{th,2}$.
These threshold voltages 
are determined in the following.
\begin{figure}[t]
\vspace*{-0.4cm}
\begin{center}
\includegraphics[width=7.5cm]{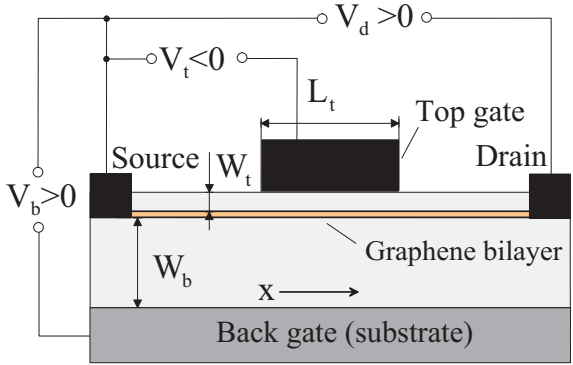}
\caption{~Schematic view of the GBL-FET structure  
}
\end{center}
\end{figure}

When $ V_{th,2} < V_{th,1} < V_t $, the top of the conduction 
band in the gated section is below the Fermi level (Fig.~2a).
In this case, an n$^+$-n-n$^+$ structure is formed in the GBL channel.
At $ V_{th,2} < V_t < V_{th,1}$,
the Fermi level is between the top of the conduction band and
the bottom of the valence band in this section (Fig.~2b).
This top-gate voltage range corresponds to 
the formation of an n$^+$-i-n$^+$ structure. 
If  $V_t < V_{th,2} < V_{th,1}$, both band edges are above the Fermi level
(Fig.~2c), so that n$^+$-p and p-n$^+$ junctions are formed beneath
the edges of the top gate.
In the first  and third ranges of the top gate voltage (``a'' and ``b''
ranges), the electron and hole 
populations of  the gated section are essential.
In the second range (range ``b''), the gated section is depleted.
 In the voltage range ``a'', 
the source-drain current is  associated with a hydrodynamical
electron flow (due to effective electron-electron 
scattering) in  the gated section.
In this case,
the source-drain current
and GBL-FET characteristics
are determined by the conductivity of the gated section,  which, in turn,
is determined by the electron density and scattering 
mechanisms including the electron-electron scattering mechanism, 
and by the self-consistent electric field directed in the channel plane. 
In such a situation, different hydrodynamical models 
of the electron transport (including the drift-diffusion model) 
can be applied
(see, for instance, Refs.~\cite{23,24,25,26,27}).  
%%%%%%%%%
\begin{figure}[t]
\vspace*{-0.4cm}
\begin{center}
\includegraphics[width=7.5cm]{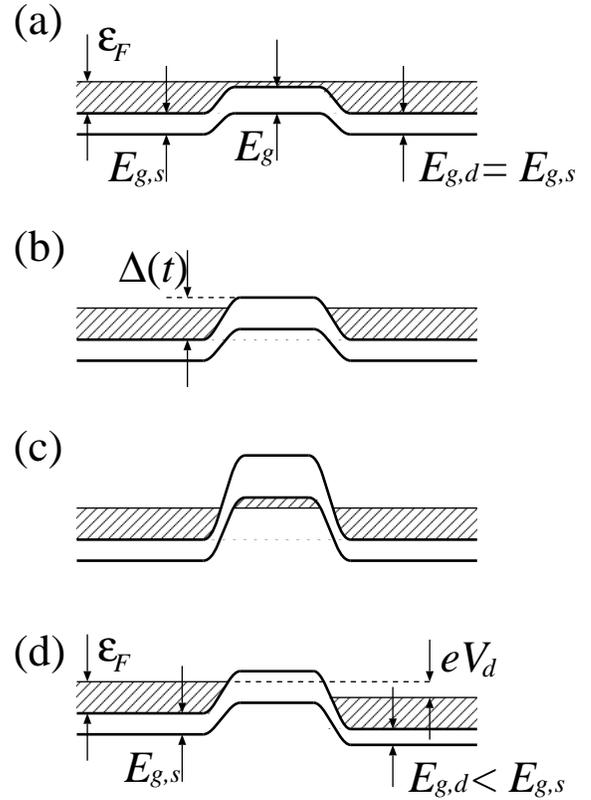}
\caption{~Band diagrams at different top 
gate bias voltages ($V_b > 0$, $V_d = 0$):
(a) $V_{th,2} < V_{th,1} < V_t$,  
(b) $V_{th,2} <  V_t < V_{th,1}$ (depleted gated section), and
(c)  $V_t < V_{th,2} < V_{th,1}$ (gated section filled with holes),
Panel (d) corresponds to $V_{th,2} <  V_t < V_{th,1}$ but  with $V_d > 0$.
}
\end{center}
\end{figure}
%%%%%%%%%%%%
If $V_{th,2} < V_t < V_{th,1}$,
considering the potential distribution in the direction
perpendicular to the GBL plane invoking the gradual channel 
approximation~\cite{28,29} and 
assuming for simplicity that
the thicknesses of the gate layers separating
the channel and the pertinent gates, $W_b$ and $W_t$, are equal to each other
$W_b = W_t = W$, we obtain
\begin{equation}\label{eq2}
\Delta_0 = - e \frac{(V_b + V_t)}{2}, 
\qquad \delta\Delta(t) = -\frac{1}{2} e\delta V(t),
\end{equation}
where $e = |e|$ is the electron charge.
In the voltage range in question, the electron system in the gated section
is not degenerate. This voltage range
as well as the range $V_t < V_{th,2} < V_{th,1}$
 correspond to the GBL-FET ``off-state''.
Similar formulas take place for the barrier height from the
drain side (with the replacement of $\Delta_0$ by $\Delta_0 + eV_d$),

In the cases when $ V_{th,2} < V_{th,1} < V_t $ or 
$V_t < V_{th,2} < V_{th,1}$,
$$
\Delta_0 = - e \frac{(V_b + V_t)}{2} 
\pm \frac{2\pi\,eW}{\kappa}\Sigma_0^{\mp},
$$
\begin{equation}\label{eq3}
 \delta\Delta(t) = - \frac{1}{2}e\delta V(t)
\pm \frac{2\pi\,eW}{\kappa}\,\delta\Sigma^{\mp}(t).
\end{equation}
Here $\Sigma_0^{\mp} + \delta\Sigma^{\mp}(t)$ are the electron
and hole densities in the gated section and $\kappa$
is the dielectric constant of the gate layers.
In the most interesting case when the electron densities
in the source and drain sections are sufficiently large, so that
the electron systems in these sections are degenerate. 
Considering this,
the height of the barrier $\Delta_0$ is given  by

\begin{equation}\label{eq4}
\Delta_0  = 
-e\frac{V_t(a_B/8W)}{[1 + (a_B/4W)]} \simeq - eV_t\biggl(\frac{a_B}{8W}\biggr), 
\end{equation}
\begin{equation}\label{eq5}
 \Delta_0  = 
-e\frac{V_t(a_B/8W) + (V_t - V_b)(d/2W)}{[1 + (a_B/4W)]}
%\simeq  e\biggl(\frac{a_B}{8W}\biggr)
%[ aV_b - (1 + a)V_t] 
\end{equation}
when  $ V_{th,2} < V_{th,1} < V_t $ and $V_t < V_{th,2} < V_{th,1}$, 
respectively.
Here $a_B = \kappa\hbar^2/me^2$ is the Bohr radius,
%$d = d_0/(1 + a)$ 
$d$
is the effective spacing between 
the graphene layers in the GBL which accounts for the screening
of the electric field between these layers~\cite{20,21}. 
This quantity is smaller than
 the real spacing between 
the graphene layers in the GBL
$d_0 \simeq 0.36$~nm.
%$a = 4d_0/a_B$.
The Bohr radius $a_B$ and parametes $a$ can be rather different
in  different materials of the gate layers.
In the cases of Si0$_2$ and Hf0$_2$
(with $k \simeq~20$~\cite{30,31}) gate
gate layers,  $a_B \simeq 4$~nm and $a_B \simeq 20$~nm, respectively.
% so that
%$a  \simeq 0.36$ and $a \simeq 0.07$, respectively.
In deriving Eqs.~(4) and (5),  
we have taken into account that in real GBL-FETs, $(a_B/8W) \ll 1$.

The Fermi energy in the source section is given by
\begin{equation}\label{eq6}
\varepsilon_F = \frac{k_BT}{[1 + (a_B/8W)]} 
\ln\biggl[
\exp\biggl(\frac{a_B}{8W}\frac{eV_b}{k_BT}\biggl) - 1\biggr], 
\end{equation}
where $k_B$ is the Boltzmann constant and $T$ is the temperature,
so that at sufficiently large back gate voltages,
\begin{equation}\label{eq7}
\varepsilon_F \simeq eV_b\frac{(a_B/8W)}{[1 + (a_B/8W)]}
\simeq eV_b\biggl(\frac{a_B}{8W}\biggr)  \gg k_BT.
\end{equation} 
Here
 we have considered that the electron density in the source section
$ \Sigma_s^{-} = \kappa\,V_b/4\pi\,eW$ 
(the electron density in the drain section of the channel
is approximately equal to $ \Sigma_d^{-} = \kappa\,(V_b - V_d)/4\pi\,eW$). 
Comparing Eqs.~(2), (4), and (5), one can see that
the height of the barrier $\Delta_0$ increases with increasing
absolute value of the top-gate voltage rather slow
in the voltage ranges ``a'' and ``c'' in contrast to
its steep increase in the voltage range ``b''.

Since the energy gaps in GBLs $E_{g,s}$, $E_{g}$, and $E_{g,d}$
depend on the local transverse electric 
field~\cite{20,21,22},
so that they are  different in different sections of the channel
depending on the bias voltages:
\begin{equation}\label{eq8}
E_{g,s} = \frac{edV_b}{2W},\,
E_{g} = \frac{ed_0(V_b - V_t)}{2W},\,
E_{g,d} = \frac{ed(V_b - V_d)}{2W}.
\end{equation}
One can see that at $V_t < 0$ and $V_d > 0$, 
one obtains $E_g > E_{g,s}\geq E_{g,d}$.

The threshold voltages $V_{th,1}$ and $V_{th,2}$ are determined
by the conditions 
$\Delta_0 = \varepsilon_F$ and  $\Delta_0 = \varepsilon_F + E_g$,respectively.
 The latter implies that the Fermi energy of holes in the gated section 
$\varepsilon_F^{(hole)} = \Delta_0 - \varepsilon_F - E_g = 0$.
As a result, the threshold voltages are given by
\begin{equation}\label{eq9}
V_{th,1} \simeq - V_b\biggl(1 + \frac{a_B}{4W}\biggr),\qquad
V_{th,2}\simeq  - V_b 
\biggl(1 + \frac{a_B}{4W} + \frac{d_0}{W}\biggr).
\end{equation}
Since one can assume that $d \ll W$, the threshold voltages are close
to each other: $|V_{th,1}| \lesssim |V_{th,2}|$
with $|V_{th,2} -  V_{th,1}| \simeq (2d_0/W)V_b \gtrsim 4E_{g,s}/e$.
The values of the energy gap 
in the gated section at the threshold 
top gate voltages are given by
\begin{equation}\label{eq10}
E_g\biggr|_{V_t = V_{th,1}} \lesssim E_g\biggr|_{V_t = V_{th,2}}
\simeq  \frac{ed_0V_b}{W} \simeq  2E_{g,s}\biggl(\frac{d_0}{d}\biggr).
\end{equation}
%

%%%%%%%
In the following we restrict our consideration by
 the situations when the height of
the barrier for electrons in the gated section is sufficiently 
large (so that $\Delta_0 > \varepsilon_F$), 
which corresponds to the band diagrams shown in Figs.~2(b) and  2(c).
% i.e., to the GBL-FET ``off-states''.

%%%%%%%%%%%%%%%%%%%%%%%%%%%

\section{Boltzmann kinetic equation and its solutions}

The quasi-classical Boltzmann kinetic equation governing the electron distribution
function
$f_{{\bf p}} = f_{{\bf p}}(x,t)$ in the section of the channel covered by the top gate (gated section)
can be presented as

\begin{equation}\label{eq11}
\frac{\partial\, f_{{\bf p}}}{\partial t} + v_x\frac{\partial\, f_{{\bf p}} }
{\partial \,x}
= \int d^2{\bf q} w(q)(f_{{\bf p} + {\bf q}} - f_{{\bf p}})
\delta (\varepsilon_{{\bf p} + {\bf q}} - \varepsilon_{\bf p}).
\end{equation}
Here, taking into account that the electron (and hole)
dispersion relation at the energies close to the
bottom of the conduction band is virtually
parabolic with the effective mass $m$ ($m \simeq 4\times10^{-29}$~g),
for the energy of electron with momentum ${\bf p} = (p_x, p_y)$
we put $\varepsilon_{\bf p} = p^2/2m = \varepsilon$,
$v_x = p_x/m = p\cos\theta/m$, where $\cos\theta = p_x/p$
(the $x$-axis and the $y$-axis
are directed in the GBL plane) and
$w(q)$ is the probability of the electron scattering on disorder
and acoustic phonons  with 
the variation of the electron momentum by quantity $q$.

The density of the electron (thermionic) current, $J = J(x,t)$, 
in the gated section
of the channel (per unit length in the $y$-direction)
can be calculated using the following formula:
\begin{equation}\label{eq12}
J = \frac{4e}{(2\pi\hbar)^2}\int\,d^2{\bf p}\,v_x\,f_{{\bf p}},
\end{equation}
where $\hbar$ is the reduced Planck constant.

Disregarding the electron-electron collisions 
in the gated section of the channel
(due a low electron density
in this section in contrast to the source and drain sections
where   the electron-electron collisions
are essential), we consider two limiting cases:
ballistic transport of electrons across the gated section
and strongly collisional electron transport.

\subsection{Ballistic electron transport}

If $\delta V(t) = \delta V_{\omega}\,e^{-i\omega\,t}$,
where $\delta V^{\omega} \ll |V_t|$ and $\omega$ are the amplitude and
frequency of the ac signal, then the electron distribution function can be searched as 
$f_{\bf p} = F_{0} + \delta F_{\omega}(x)\,e^{- i\omega\,t}$
and $\Delta = \Delta_0 + \delta\Delta_{\omega}\,e^{- i\omega\,t}$.
Assuming that $eV_d \gg k_BT$ and solving Eq.~(11) 
with the boundary conditions 
\begin{equation}\label{eq13}
f_{\bf p}\biggr|_{p_x \geq 0, x = 0} = 
\exp\biggl[\frac{\varepsilon_F - \Delta(t) -  \varepsilon}{k_BT}\biggr], 
\qquad f_{\bf p}\biggr|_{p_x \leq 0, x = L_t} \simeq 0,
\end{equation}
where $L_t$ is the top gate length,
we obtain
\begin{equation}\label{eq14}
 F_{0} \simeq 
\exp\biggl(\frac{\varepsilon_F - \Delta_0 -  \varepsilon}{k_BT}\biggr)\,
\Theta(p_x),
\end{equation}
$$
\delta F_{\omega}(x) = 
\exp\biggl(\frac{\varepsilon_F - \Delta_0 -  \varepsilon}{k_BT}
+ i\omega\,\sqrt{\frac{m}{2\varepsilon}}\frac{x}{\cos\Theta}\biggr)
$$
\begin{equation}\label{eq15}
\times
\biggl(- \frac{\delta\,\Delta_{\omega}}{k_BT}\biggl)\,\Theta(p_x).
\end{equation}
Here, $\Theta(p_x)$ is the unity step function.
The first boundary condition given by Eq.~(14) corresponds to quasi-equilibrium
electron distribution in the source section of the channel and
the injection of electrons with the kinetic energy exceeding 
the barrier height $\Delta(t)$ from the source section to the gated section
(at $x = 0$).
The injection of electrons from the drain source to the gated section
(at $x = L_t$)
is negleted due to $eV_d \gg k_BT$; this inequality leads to
rather  high barrier near the drain edge of the gated section. 
The presence of the unity step fuction $\Theta(p_x)$ in Eqs.~(16) and (17)
reflects the fact that there are no electrons propagating backwards
due to the absence of the electron scattering in the gated section. 

Using Eqs.~(12), (14), and (15), we arrive at the following formulas for the
dc and ac components, $J_0$ and $\delta\,J_{\omega}$,
 of the current at the drain edge of the gated section (i.e., at $x = L_t$)

\begin{equation}\label{eq16}
J_0 = e\frac{\sqrt{2m}(k_BT)^{3/2}}{\pi^{3/2}\hbar^2}
\exp\biggl(\frac{\varepsilon_F - \Delta_0}{k_BT}\biggr) = J_0^{B},
\end{equation}

\begin{equation}\label{eq17}
\frac{\delta\,J_{\omega}}{J_0} = 
\biggl(- \frac{\delta\,\Delta_{\omega}}{k_BT}\biggr)
\int_0^{\infty}d\xi\,\sqrt{\xi}\,e^{- \xi}\,
{\cal F}_{\omega}(\xi).
%\simeq \biggl(- \frac{\delta\,\Delta_{\omega}}{k_BT}\biggr)\sqrt{\frac{\pi}{%2}}{\cal F}_{\omega}(1/2).
\end{equation}
Here 
$$
{\cal F}_{\omega}(\xi) 
= 
 \frac{2}{\sqrt{\pi}}\int_0^{1}dy\,
\exp\biggl(i\frac{\omega\tau}{\sqrt{\xi}\sqrt{1 - y^2}}\biggr)
$$
$$
\simeq 2\frac{\xi^{1/4}}{\sqrt{\omega\tau}}\exp\biggl(i\frac{\omega\tau}
{\sqrt{\xi}}\biggr)
\biggl[C\biggl(\sqrt{\frac{\omega\tau}{2\sqrt{\xi}}}\biggr) + iS
\biggl(\sqrt{\frac{\omega\tau}{2\sqrt{\xi}}}\biggr)\biggr],
$$
where
$\tau = L_t\sqrt{m/2k_BT}$ is the effective
 ballistic transit time across the gated section of electrons
with the thermal velocity $v_T = \sqrt{2k_BT/m}$ and
 $C(x)$ and $S(x)$ are Frenel's cosine and sine functions.
At $\omega\tau \gg 1$,
$$
{\cal F}_{\omega}(\xi) \simeq \frac{\sqrt{2}\xi^{1/4}}{\sqrt{\omega\tau}}
\exp\biggl(i\frac{\omega\tau}{\sqrt{\xi}} + i\frac{\pi}{4}\biggr).
$$
At $\omega\tau \ll 1$, ${\cal F}_{\omega}(\xi)$ tends to unity.

%%%%%%%%%%%%%%%%%%%%%%%%%%%%%%%%%%%%%
\subsection{Collisional electron transport}

In the case of strongly collisional electron transport,
the distribution function in the gated section is close to isotropic
and it can be
 searched in the form 
\begin{equation}\label{eq18}
f_{\bf p} = F +  g\,\cos\theta.
\end{equation}
Here %$\cos \theta = p_x/p$, 
 $F = F(\varepsilon, x,t)$ is  the symmetrical part of
the electron distribution function 
(which is  generally not the equilibrium function). 
The second term in Eq.~(18) presents the asymmetric
part of the distribution function with
 $g = g(\varepsilon, x,t)$.
Similar approach was used  for the calculation 
of characteristics of heterojunction bipolar transistors~\cite{32}
(see also Refs.~\cite{19,33}).
As a result, after the averaging of Eq.~(11)
over the angle $\theta$, one can arrive at the following coupled
equations:

\begin{equation}\label{eq19}
\frac{\partial F}{\partial t} = 
- \sqrt{\frac{\varepsilon}{2m}}\frac{\partial g}{\partial x},
\qquad
\frac{\partial gF}{\partial t} + \nu\,g = 
- \sqrt{\frac{2\varepsilon}{m}}\frac{\partial F}{\partial x}.
\end{equation}
Here in the case of $w(q) = w = const$, which corresponds
to the scattering of electrons on short-range defects,  
$\nu  = m w/2$.
Equations~(19)  are reduced to the following equation
for function $F$:
\begin{equation}\label{eq20}
\frac{\partial}{\partial t}\biggl(\frac{\partial\,F_{\varepsilon}}
{\partial\,t} 
+ \nu\,F\biggr) 
= 
\frac{\varepsilon}{m}\frac{\partial^2 F }{\partial\, x^2}.
\end{equation}
In the most interesting case when $eV_d \gg k_BT$,
the boundary conditions for Eq.~(20)
at $x = 0$  and $x = L_t$, can be adopted in the following form:
\begin{equation}\label{eq21}
F\biggr|_{x = 0} =
\exp\biggl[\frac{\varepsilon_F - \Delta(t) - \varepsilon}{k_BT}\biggr],\qquad
 F\biggr|_{x = L_t} \simeq 0.
\end{equation}
The boundary condition under consideration imply that at $x = 0$
there is the electron injection from the source section of the channel,
whereas at $x = L_t$ an effective extraction of the electrons
into the drain section occurs due to a strong pulling dc electric field.
Due to a strong electron scattering a significant portion
of the injected electrons returns back to the source section. 

Setting as above 
$\delta V(t) = \delta V^{\omega}\,e^{-i\omega\,t}$ and, hence,
$F = F_0 + \delta F_{\omega}\,e^{-i\omega\,t}$ and
 $g = F_0 + \delta g_{\omega}\,e^{-i\omega\,t}$, we obtain
\begin{equation}\label{eq22}
\frac{d^2 F_0 }{d\, x^2} = 0,
\end{equation}
\begin{equation}\label{eq23}
\frac{d^2 \delta\,F_{\omega} }{d\, x^2} 
- \frac{m\omega(\omega + i\nu_p )}{\varepsilon} \delta\,F_{\omega}= 0,
\end{equation}
and  arrive at
\begin{equation}\label{eq24}
F_0 = \exp\biggl[\frac{\varepsilon_F - \Delta_0 - \varepsilon}{k_BT}\biggr]
\biggl(1 - \frac{x}{L_t}\biggr),
\end{equation}
\begin{equation}\label{eq25}
\delta F_{\omega} = 
\exp\biggl[\frac{\varepsilon_F - \Delta_0 - \varepsilon}{k_BT}\biggr]
\frac{\sinh[\alpha_{\omega}(x - L_t)]}{\sinh(\alpha_{\omega}L_t)}
\biggl(\frac{\delta\,\Delta_{\omega}}{k_BT}\biggr),
\end{equation}
where $\alpha_{\omega} = \sqrt{m\,\omega(\omega + i\nu)/\varepsilon}$.
Considering Eqs.~(19), (24), and (25) , we obtain
$$
g_0 = - \frac{1}{\nu}\sqrt{\frac{2\varepsilon}{m}}
\frac{\partial F_0}{\partial x}
$$
\begin{equation}\label{eq26}
= \frac{1}{\nu\,L_t}\sqrt{\frac{2\varepsilon}{m}}
\exp\biggl[\frac{\varepsilon_F - \Delta_0 - \varepsilon}{k_BT}\biggr],
\end{equation}

$$
\delta\,g_{\omega} = - \frac{i}{(\omega + i\nu)}
\sqrt{\frac{2\varepsilon}{m}}
\frac{\partial\, \delta F_{\omega}}{\partial x}
$$
$$
= - i
\exp\biggl[\frac{\varepsilon_F - \Delta_0 - \varepsilon}{k_BT}\biggr]
\sqrt{\frac{2\omega}{(\omega + i\nu)}}
$$
\begin{equation}\label{eq27}
\times\frac{\cosh[\alpha_{\omega}(x - L_t)]}{\sinh(\alpha_{\omega}L_t)}
\biggl(\frac{\delta\,\Delta_{\omega}}{k_BT}\biggr).
\end{equation}
After that, using Eqs.~(12), (26), and (27),
we arrive at the following formulas for
$J^0$ and $\delta J^{\omega}$ (at $x = L_t$):
\begin{equation}\label{eq28}
J_0 = e\frac{2(k_BT)^2}{\pi\hbar^2L_t\nu}
\exp\biggl(\frac{\varepsilon_F - \Delta_0}{k_BT}\biggr) = J_0^{C},
\end{equation}

\begin{equation}\label{eq29}
\frac{\delta\,J_{\omega}}{J_0} = 
\biggl(- \frac{\delta\,\Delta_{\omega}}{k_BT}\biggr)
\int_0^{\infty}d\xi\,\sqrt{\xi}\,e^{-\xi}
{\cal H}_{\omega}(\xi)
.
\end{equation}
Here
$$
{\cal H}_{\omega}(\xi) = \frac{i\,\tau\nu}
{\sinh\sqrt{[2\omega(\omega + i\nu)\tau^2/\xi]}}
\sqrt{\frac{2\omega}{\omega + i\nu}}.
$$
According to Eq.~(28),
$J_0^{C} \propto 1/L_t\nu$.
One needs to stress that the collisional case under consideration
corresponds actually to
$\nu\tau \gg 1$.
In the frequency range $\omega \ll \nu/\tau^2$, 
${\cal H}_{\omega}(\xi) \simeq \sqrt{\xi}$.
At  $\omega \gg \nu/\tau^2$, one obtains 
$$
{\cal H}_{\omega}(\xi) \simeq 2(1 + i)\sqrt{\omega\tau^2\nu}
\exp\biggl[- \frac{(1 + i)\sqrt{\omega\tau^2\nu}}{\sqrt{\xi}}\biggr].
$$

\section{GBL-FET dc transconductance}

Equations~(16) and (28) provide the dependences of the source-drain
dc current $J_0$ as a function of the device structural parameters, 
temperature, and 
back- and top-gate voltages for GBL-FETs with ballistic and collisional electron
transport, respectively (in the limit $eV_d \gg k_BT$).
Using Eq.~(16), one can find
the dc transconductance $G_0 = (\partial J_0/\partial V_t)|_{V_b}$
 of a GBL-FET with the ballistic electron transport:
\begin{equation}\label{eq30}
G_0^{B}  =  e^2\frac{\sqrt{2mk_BT}}{\pi^{3/2}\hbar^2}
\exp\biggl(\frac{\varepsilon_F - \Delta_0}{k_BT}\biggr)
= \frac{eJ_0^{B}}{2k_BT}
\end{equation}
when  $V_{th,2} < V_t< V_{th,1}$,
and 

\begin{equation}\label{eq31}
G_0^{B}  =  e^2\frac{\sqrt{2mk_BT}}{\pi^{3/2}\hbar^2}\exp\biggl(\frac{\varepsilon_F - \Delta_0}{k_BT}\biggr){\cal R}_0
= \frac{eJ_0^B}{2k_BT}\,{\cal R}_0
\end{equation}
when $V_t < V_{th,2} < V_{th,1}$.
Here
\begin{equation}\label{eq32}
{\cal R}_0 \simeq \biggl(\frac{a_B}{4W}\biggr).
%\biggl[\frac{1 + (4d/a_B)}
%{1 + (a_B/4W)}\biggr] \simeq 
%(1 + a)\biggl(\frac{a_B}{4W}\biggr).
\end{equation}

Similarly, using Eq.~(28), we obtain the following formulas for
the GBL-FET transconductance in the case of collisional electron transport:
\begin{equation}\label{eq33}
G_0^{C}  =  e^2\frac{k_BT}{\pi\hbar^2L_t\nu}
\exp\biggl(\frac{\varepsilon_F - \Delta_0}{k_BT}\biggr)
= \frac{eJ_0^{C}}{2k_BT}
\end{equation}
when  $V_{th,2} < V_t < V_{th,1}$
and 
\begin{equation}\label{eq34}
G_0^{C}  =   e^2\frac{k_BT}{\pi\hbar^2L_t\nu}
\exp\biggl(\frac{\varepsilon_F - \Delta_0}{k_BT}\biggr)\,{\cal R}_0
= \frac{eJ_0^{C}}{2k_BT}\,{\cal R}_0
\end{equation}
when $V_t < V_{th,2} < V_{th,1}$.

As follows from the comparison of Eq.~(30) with Eq.~(31)
and Eq.~(33) with Eq.~(34), the GBL-FET dc transconductance
in the top gate voltage range   $V_{th,2} < V_t < V_{th,1}$ (in the range ``b'')
might be much larger than that when $V_t < V_{th,2} < V_{th,1}$
(in the range ``c'') since  $a_B \ll 8W$. This is due to relatively
slow increase in $\Delta_0$ with increasing $|V_t|$
when the hole density in the gated section becomes essential.

%%%%%%
The voltage dependences of the dc transconductance can be obtained
using Eqs.~(30), (31), (33), and (34) and  invoking Eqs.~(2), (4), and (6).
In particular, in a rather narrow voltage range $V_{th,2} < V_t< V_{th,1}$,
one obtains
\begin{equation}\label{eq35}
G_0^{C} =  \frac{\sqrt{\pi}}{\nu\tau}\,G_0^{B} \propto 
\exp\biggl[\frac{e(V_t - V_{th,1})}{2k_BT}\biggr].
\end{equation}
At sufficiently large absolute values of the top-gate voltage
when  $ V_t < V_{th,2} < V_{th,1}$, the transconductance vs voltage 
dependence is given by
%
%$$
%G_0^{C}  =  \frac{\sqrt{\pi}}{\nu\tau}\,G_0^{B}\propto 
%(1 + a)\biggl(\frac{a_B}{4W}\biggr)\exp\biggl\{\frac{e[V_b(1 - a) + V_t(1 + a)]}
%{k_BT}\biggl(\frac{a_B}{8W}\biggr)
%\biggr\} 
%$$
$$
G_0^{C}  =  
\frac{\sqrt{\pi}}{\nu\tau}\,G_0^{B}\propto \biggl(\frac{a_B}{4W}\biggr)
\exp\biggl[\frac{e(V_t - V_b)}{k_BT}\biggl(\frac{d_0}{2W}\biggr)
\biggr]
$$
\begin{equation}\label{eq36}
\times\exp\biggl[\frac{e(V_t - V_{th,2})}{k_BT}\biggl(\frac{a_B}{8W}\biggr) 
\biggr].
\end{equation}
%

%%%%%%%%%%%%%%%%%%%%%%%%%%%%%%%%
\section{GBL-FET ac transconductance}

According to the Shockley-Ramo theorem~\cite{34,35},
the source-drain ac current is equal to the ac current induced 
in the highly conducting
 quasi-neutral portion of the drain section of the channel
and in the drain contact
by
the electrons injected from the gated section. This current is determined
by the injected ac current given by Eq.~(16) or Eq.~(28) as well as
the electron transit-time effects in the depleted portion of the drain
section. However, if $\omega\tau_d \ll 1$, where the $\tau_d$
is the electron transit time in depleted region in question,
the induced ac  current is very close to the injected  
ac current~\cite{19,36,37}.
Since, at moderate drain voltages, the length  of depleted portion of the drain
section $L_d$ can usually be shorter than the top gate length $L_t$,
in the most practical range of the signal frequencies
$\omega \lesssim \tau^{-1}$, one can assume that $\tau_d \ll \tau$ and, hence,
 $\omega\tau_d \ll 1$.
Considering this and 
using Eqs.~(17) and (28), the GBL-FET ac transconductance $G_{\omega} = 
(\partial \delta\,J_{\omega}/\partial \delta\,V_{\omega})|_{V_b}$
at different electron transport conditions can be presented
as 
\begin{equation}\label{eq37}
G_{\omega}^B = \frac{J_0^B}{k_BT}
\biggl(- \frac{\partial\delta\Delta_{\omega}}
{\partial \delta\,V_{\omega}}\biggr|_{V_b}\biggr)
\int_0^{\infty}d\xi\,\sqrt{\xi}\,e^{- \xi}\,
{\cal F}_{\omega}(\xi),
\end{equation}

\begin{equation}\label{eq38}
G_{\omega}^C =   \frac{J_0^C}{k_BT}
\biggl(- \frac{\partial \delta\Delta_{\omega}}
{\partial \delta\,V_{\omega}}\biggr|_{V_b}\biggr)  
\int_0^{\infty}d\xi\,\sqrt{\xi}\,e^{-\xi}{\cal H}_{\omega}(\xi),
\end{equation}
respectively.

In the range of  gate voltages $V_{th,2} < V_t< V_{th,1}$
(range ``b''), Eq.~(2) yields 
\begin{equation}\label{eq39}
\frac{\partial\,\delta\Delta_{\omega}}{\partial\,\delta V\omega}\biggr|_{V_d} 
= - \frac{e}{2}.
\end{equation}
In this case, Eqs.~(37) and (38) result in   
\begin{equation}\label{eq40}
G_{\omega}^B = \frac{eJ_0^B}{2k_BT}
\int_0^{\infty}d\xi\,\sqrt{\xi}\,e^{- \xi}\,
{\cal F}_{\omega}(\xi),
\end{equation}

\begin{equation}\label{eq41}
G_{\omega}^C =   \frac{eJ_0^C}{2k_BT}
\int_0^{\infty}d\xi\,\sqrt{\xi}\,e^{-\xi}{\cal H}_{\omega}(\xi).
\end{equation}

As follows from Eqs.~(40) and (41),
the characteristic frequencies of the ac transconductance roll-off are
$1/\tau$ and $\nu/\tau^2$ in the case of the ballistic 
and collisional electron
transport, respectively, i.e.,  the inverse times of the ballistic and
diffusive transit across the gated section of the channel. 
Indeed, the quantity  $\nu/\tau^2$ can be presented as $D/L_t^2$,
where $D$ is the electron diffusion coefficient.

\begin{figure}[t]
\vspace*{-0.4cm}
\begin{center}
\includegraphics[width=7.5cm]{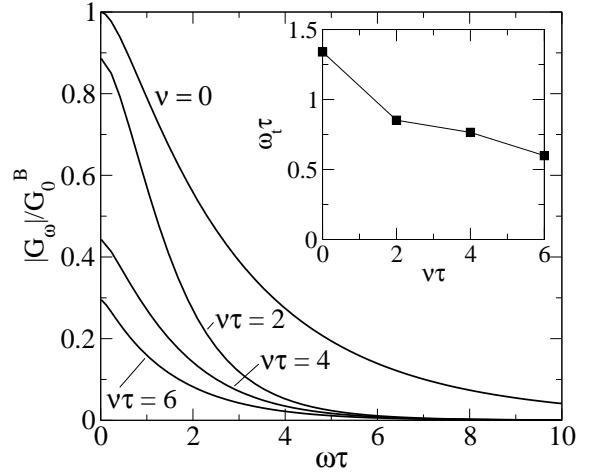}
\caption{~The ac transconductance (normalized by $G_0^B$) versus $\omega\tau$
calculated for GBL-FETs with ballistic
($\nu = 0$) and collisional electron transport
($\nu\tau = 2, 4,$ and 6) at the top gate voltage in range ``b'', i.e.,
 $V_{th,2} <  V_t < V_{th,1}$.
}
\end{center}
\end{figure}

The situation becomes more complex in the range 
of the top gate bias voltages $V_t < V_{th,2} < V_{th,1}$ (range ``c''). 
As follows from Eq.~(4) in this voltage range, 
the quantity $\delta\,\Delta_{\omega}$ is determined
not only by the ac voltage $\delta\,V_{\omega}$ but
also by the ac component of the hole density in the gated
section $\delta\Sigma^{+}_{\omega}$.
Moreover, at sufficiently high signal frequencies,
the hole system in the gated section can not manage to follow
the variation of the ac voltage.
Taking into account the dynamic response of the hole system
(see Appendix A), instead of Eq.~(39) one can obtain
 
\begin{equation}\label{eq42}
\frac{\partial\,\delta\Delta_{\omega}}{\partial\,\delta V_\omega}\biggr|_{V_d} 
= 
- \frac{e}{2}{\cal R}_{\omega}. 
\end{equation}
Here (see Appendix A)
$$
{\cal R}_{\omega} = 
\biggl(\frac{a_B}{4W}\biggr)
\biggl\{\biggl(\frac{a_B}{4W}\biggr)  + \displaystyle
\frac{1}  
{1  - i\omega\tau_r}
\biggr\}^{-1} 
$$
\begin{equation}\label{eq43}
= {\cal R}_0 
\biggl(\frac{1  - i\omega\tau_r}{1  - i\omega\tau_r{\cal R}_0}\biggr),
\end{equation}
where $\tau_{r}$ is the time
of the gated section recharging associated with changing
of the hole density due to the tunneling or/and generation-recombination
processes. Generally,  $\tau_{r}$ depends on the top gate length $L_t$.
%This is because $\tau_{RC} \propto L_t$, while $\tau_g$ is independent of
%$L_t$.

Accounting for  Eq.~(42), 
we arrive at the following formulas for the GBL-FET  ac transconductance
when $V_t < V_{th,2} < V_{th,1}$:  
\begin{equation}\label{eq44}
G_{\omega}^B = \frac{eJ_0^B}{2k_BT}{\cal R}_{\omega}
\int_0^{\infty}d\xi\,\sqrt{\xi}\,e^{- \xi}\,
{\cal F}_{\omega}(\xi),
\end{equation}

\begin{equation}\label{eq45}
G_{\omega}^C =   \frac{eJ_0^C}{2k_BT}{\cal R}_{\omega}
\int_0^{\infty}d\xi\,\sqrt{\xi}\,e^{-\xi}{\cal H}_{\omega}(\xi).
\end{equation}
If $\omega \gg \tau_{r}^{-1}$, one obtains
${\cal R}_{\omega} \simeq 1$ and  the ac transconductances in
both ``b'' and ``c'' ranges of the top gate voltage are close to each other
(compare Eqs.~(40) and (41) with Eqs.~(42) and (44)).
However, at low signal frequencies ( $\omega \gg \tau_{r}^{-1}$), 
the ac transconductance
given by Eq.~(44) or (45) for the voltage range ``c''
 are markedly smaller
than those given by Eqs.~(40) or (41) valid in the voltage range ``b''.

%%%%%%%%%%%
\begin{figure}[t]
\vspace*{-0.4cm}
\begin{center}
\includegraphics[width=7.0cm]{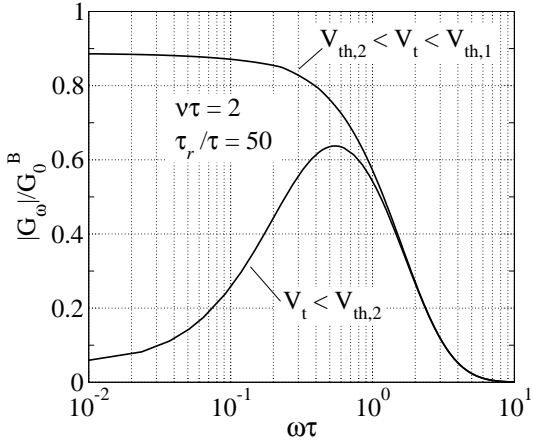}
\caption{The ac transconductance (normalized by $G_0^B$) versus $\omega\tau$
calculated for GBL-FETs with collisional electron transport
($\nu\tau = 2$)
at the top voltage in range ``c''
( $V_t < V_{th,2} < V_{th,1}$) at $\tau_r/\tau = 50$.
}
\end{center}
\end{figure}

\section{Analysis of the results  and discussion}
Comparing $G_0^{B}$ and $G_0^{C}$ given by Eqs.~(30) and (35), we obtain
$G_0^{C}/G_0^{B} = \sqrt{\pi}/\nu\tau$. This implies that
the above ratio markedly 
decreases with increasing collision frequency (with decreasing electron mobility) and the top gate length, i.e., with the departure from the ballistic transport.
%%%%%%%%%%

As shown above, the dc current steeply drops in a narrow
top-gate  voltage range $V_{th,2} < V_t < V_{th,1}$.
Inded, the ratio $J_0^B|_{V_t = V_{th,2}}/ J_0^B|_{V_t = V_{th,1}} \simeq
\exp[ - (ed_0V_b/Wk_BT)]$.  Setting $W = 5$~nm, $T = 300$~K,
 and $V_b = 1 - 2$~V,
we find $J_0^B|_{V_t = V_{th,2}}/ J_0^B|_{V_t = V_{th,1}} \simeq
3\times10^{-3} - 6\times10^{-2}$.
The estimate of the  dc current at $V_t = V_{th,2} \simeq - V_b$
(which might be interesting for the GBL-FET applications in digital large scale
circuits) 
with $W = 5$~nm, $T = 300$~K,
 and $V_b = 1 - 2$~V yields 
$J_0^B \simeq 1\times10^{-3} - 2\times10^{-2}$~A/cm.
At $T = 77$~K and $V_b = 1$~V, one obtains $J_0^B \simeq 7\times10^{-7}$~A/cm.
In the case of a GBL-FET with the width $H = 1~\mu$m, the latter corresponds to
the characteristic value of the ``off- current'' 
$J_0^BH \simeq 70$~pA. Similar values can be obtained at $T = 300$~K
when $-V_t = V_b \simeq 4.6$~V.

As follows from Eq.~(35), the GBL dc transconductance 
in the range of the top gate voltages  $V_{th,2} < V_t < V_{th,1}$
is particularly
 large when $V_t \lesssim V_{th,1}$. This is due to a sharp voltage-
sensitivity
of the dc current and its relatively high values  at such voltages.
Indeed,  using Eq.~(28)  at $T = 300$~K and $V_t \lesssim V_{th,1}$,
we obtain $G_0^B \lesssim 2500 $~mS/mm.
In GBL-FETs with $W_t \ll W_b$, the dc transconductance can be even larger.

The pre-exponential factor in the right-hand side of Eq.~(36)
is proportional to a small parameter 
$(a_B/8W)$.
The argument of the exponential function in this equation
 comprises small parameters
$(a_B/8W)$ and $(d_0/2W)$. This implies that the dc transconductance in the
voltage range  $ V_t < V_{th,2} < V_{th,1}$ described by Eq.~(36) is 
relatively small and is a faily
weak function of the top-gate voltage. As follows from Eqs.~(35) and (36),
the ratio of the dc transconductance at $V_t \lesssim V_{th,2}$ to that at  
$V_t \lesssim V_{th,1}$ is equal approximately to the following small
value: $(a_B/4W)
\exp[ed_0(V_{th,2} - V_b)/2Wk_BT] = (a_B/4W)
\exp( - 2eE_{g,s}d_0/dk_BT)$; it is smaller then the ratio of the dc currents
by parameter $(a_B/4W)$.

Figure~3 shows the  ac transconductance $|G_{\omega}|$ normalized by the dc
transconductance
at the ballistic transport $G_0^B$
as a function of the normalized signal frequency $\omega\tau$ 
calculated for GBL-FETs with both ballistic and collisional
electron transport. It is assumed that the top gate voltage
is in the range   $V_{th,2} < V_t <V_{th,1}$.
The inset in Fig.~3 shows the dependence of the normalized
threshold frequency
$\omega_t\tau$ on $\nu\tau \propto \nu\,L_t$. The threshold frequency
is defined as that at which $|G_{\omega}|/G_0 = 1/\sqrt{2}$.
One can see that  $|G_{\omega}|$ pronouncedly decreases with increasing
collision frequency. However, as seen from the inset in Fig.~3,
 the decrease in $\omega_t$ with increasing
$\nu$ is markedly slower: the ratio of $\omega_t$ at $\nu\tau = 2$
and   $\omega_t$ at $\nu\tau = 6$ (i.e., three times larger) 
is approximately equal to 1.42.
Setting $L_t = 100 - 500$~nm and $T = 300$~K, for the threshold
frequency $f_t = \omega_t/2\pi$ at the ballistic transport we obtain 
$f_t^B \simeq 
0.485 - 0.97$~THz.
To realize the near 
ballistic regime of the electron transport ($\nu\tau \ll 1$) 
in GBL-FETs with such
gate lengths, the electron mobility 
$\mu > (1 - 5)\times10^4$~cm$^2$/Vs is required. The possibility of
the latter mobilities at room temperatures was discussed recently
(see, for instance, Ref.~\cite{6}).
At a shorter top gate, $L_t = 75$~nm, one obtains
$f_t^B \simeq 1.29$~THz. 
%%%%%%%%%%%%%
In the case of a GBL-FET with  relatively long top gate
and moderate mobility ($L_t = 500$~nm and
$\mu = 2\times10^4$~cm$^2$/Vs)   when the effect of scattering
is strong ($\nu \simeq 2\times10^{12}$~s$^{-1}$
and $\nu\tau \simeq 2.7$), we obtain $f_t^C \simeq 94$~GHz.

In Fig.~4, 
the similar dependences  
calculated for a GBL-FET with collisional electron transport
at  $V_{th,2} < V_t <V_{th,1}$
and at  $V_t < V_{th,2} < V_{th,1}$ are demonstrated.
Since in the latter top-gate voltage range 
the electron scattering on
holes accumulated in the gated section can be  strong,
so that the realization of the ballistic transport
at such top-gate voltages might be problematic,
only the dependences corresponding to the collisional electron 
transport are shown.
As follows from Eqs.~(44) and (45) and seen from Fig.~4,
the ac transconductance at the top gate
voltages $V_t < V_{th,2} < V_{th,1}$ is fairly small at low frequencies
$\omega \lesssim \tau_r^{-1}$ being close to the dc transconductance
(due to a smallness of parameter
${\cal R} \simeq a_B/4W$), whereas it becomes much larger
in the intermediate frequency range 
$\tau_r^{-1} \ll \omega \lesssim \tau^{-1}$ (or
$\tau_r^{-1} \ll \omega \lesssim \nu\tau^{-2}$).
This is due to the effect of holes in the gated section.
Owing to this effect, the low frequency noises can be effectively
suppressed.

Using the above results  for GBL-FETs
and those obtained previously~\cite{19} for GNR-FETs,
one can compare the GBL and GNR characteristics.
In particular, considering  the expressions for $J_0^B$ found for GBL-FETs
and GNR-FETs, we obtain $J_0^B/J_0^{B,GNR} =  1/2$.
For the case of collisional transport, one obtains $J_0^C/J_0^{C,GNR} \sim 1$.
As a result, the GBL-FET and GNR-FET  dc transconductances
are close to each other.
The ratio of the GBL-FET and GNR-FET ac transconductances
at high signal frequencies is 
$G_{\omega}^B/G_{\omega}^{B,GNR} \propto 
1/\sqrt{\omega\tau}$,
i.e., the GBL-FET ac transconductance  falls more
steeply with increasing $\omega$ than the GNR-FET ac transconductance.

%%%%%%%%%!!!!!!!!!!
The GBL-FET dc and ac characteristics obtained are valid if the 
interband tunneling source-drain current 
through the $n^+-p$ and $p-n^+$ junctions beneath the edges
of the top gate are small in comparison with the thermionic current created by
the electrons overcoming the potential barrier in the gated section.
Such a tunneling current can be essential in the voltage range ''c''
($ V_t < V_{th,2} < V_{th,1}$) depending on the energy gap 
near the
top gates and the length of the   $n^+-p$ and $p-n^+$ junctions in question.
This implies that there is a limitation 
when the top-gate voltage is not too high $ V_t$  in comparison with
the threshold voltage $V_{th,2}$ (i.e., when $ V_t \lesssim V_{th,2}$),
so that the calculated characteristics correspond to the most interesting
voltage range
where the ac transconductance can be rather large.

\section{Conclusions}
We have presented an analytical device model for a GBL-FET.
Using this model, we have
calculated the GBL-FET dc and ac characteristics and shown that:\\
(1) The dependence of the dc current on the top gate voltage
is characterized by the existence of three voltage ranges,
corresponding to (a) the population of the gated section by electrons,
(b) the depletion of this section, and (c)  
its essential filling  with holes,
and 
determined by the top-gate threshold voltages $V_{th,1}$ and $V_{th,2}$.\\
(2) The ac current is most sensitive to the top-gate voltage $V_t$
and   the dc and ac transconductances are  large when 
$V_{th,2}< V_t \lesssim V_{th,2}$.\\
(3) The electron scattering in the gated section results in a marked
reduction in the dc and ac transconductances. However,
the threshold frequency corresponding to $|G_{\omega}|/G_0 = 1/\sqrt{2}$
decreases with increasing collision frequency relatively smoothly.\\
(4) The transient recharging of the gated section by holes (at $V_t < V_{th,2}$)
 leads to a nonmonotonic frequency dependence of the ac transconductance
with a pronounced maxima in the range of  fairly high frequencies.
This effect might be used for the optimization of GBL-FETs with
reduced  sensitivity to low frequency noises.\\
(5) The fabrication of GBLs with high electron mobility
at elevated temperatures opens up the prospects
of realization of terahertz GBL-FETs with ballistic electron transport 
operating at room temperatures and
surpassing FETs based on A$_3$B$_5$ compounds.

\section*{Acknowledgments}
The authors are grateful to Professor S.~Brazovskii
 (University Paris-Sud) and Professor E.~Sano (Hokkaido University)
for fruitful discussions and valuable information.
The work was supported by the Japan Science and 
Technology Agency, CREST,  Japan.
One of the authors (N.K.) acknowledges the support 
by the INTAS grant 7972, and by the ANR 
program in France (the project BLAN07-3-192276).

\section*{Appendix A. Dynamic response of the hole system in the gated section}
\setcounter{equation}{0}
\renewcommand{\theequation} {A\arabic{equation}}

At sufficiently large values $|V_t|$, the gated section
is essentially populated with the holes.
As follows from eq.~(4),
\begin{equation}\label{A1}
\delta\,\Delta_{\omega}
= - \frac{1}{2}e\delta V_{\omega}
-\frac{2\pi\,e^2W}{\kappa}\,\delta\Sigma^{+}_{\omega}.
\end{equation}
The ac component of the hole density $\delta\Sigma^{+}_{\omega}$
obeys the continuity equation which can be presented in the following form:
\begin{equation}\label{A2}
-i\omega \delta\,\Sigma_{\omega}^{+} = \delta G_{\omega} 
 + \frac{\delta\,J^{T}_{\omega}|_{x = 0} - 
\delta\,J^{T}_{\omega}|_{x = L_t}}{ eL_t}.
\end{equation}

Here $\delta G_{\omega}$ is the variation of the generation pf holes
 in the gated section (associated, say, with the generation of holes
by the thermal radiation~\cite{38}) 
%$\tau_g$ is the characteristic time of the generation-recombination
%processes
and $\delta\,J^{T}_{\omega}|_{x = 0}$ and $\delta\,J^{T}_{\omega}|_{x = L_t}$
are the interband tunneling ac currents near the source and drain
edges of the top gate, respectively.
For normal operation of GBL-FETs, these tunneling current should relatively
small. This is achieved  in GNL-FETs by proper choice
of the energy gap in the different sections of the channel
($E_{g,s}$, $E_g,$, and $E_{g,d}$ which should be not too small.
%Estimating the in-plane electric field ${\cal E}_{parallel}$ at $\x = 0$ and $x = L_t$
%as $E_{parallel} \sim \Delta_0/ea_B$ and considering eq.~(8),
The terms in thr right-hand side of eq.~(A2) can be presented as
\begin{equation}\label{A3}
\delta G_{\omega} = K_g(\delta \Delta_{\omega} - \delta\varepsilon_F^{+}),
\end{equation}
\begin{equation}\label{A4}
\delta\,J^{T}_{\omega}|_{x = 0} - \delta\,J^{T}_{\omega}|_{x = L_t} =
\frac{2(\delta \Delta_{\omega} - \delta\varepsilon_F^{+})}{eR_t}.
\end{equation}
Here $\delta\varepsilon_F^{+} 
\simeq (\pi\hbar^2/2m)\delta\Sigma_{\omega}^{+}$,
$K_g = 2m/\pi\hbar^2\tau_g$,
where $\tau_g$ is the characteristic time of the generation-recombination
processes in the gated section,
$R_t$ is the tunneling resistance of the p-n junctions induced by
the negative top gate voltage near the edges of the top gate:
 $ln R_t \propto E_g^{3/2}/{\cal E}_{\parallel}
\propto (V_b - V_t)^{3/2}/V_t$.
%%as follows from rough estimates $\ln R$ increases approximately as
%$\sqrt{V_t}$
From eqs.~(A1) - (A4),  taking into account the limit $\omega \rightarrow 0$
(see eqs.~(5) and (30)),
we find 
$$
\delta\,\Delta_{\omega}
= - \frac{e}{2}\delta V_{\omega} 
$$
$$
\times\biggl(\frac{a_B}{4W}\biggr)\biggl\{\biggl(\frac{a_B}{4W}\biggr)  + \displaystyle
\frac{\displaystyle\frac{1}{\tau_g} + 
\biggl(\frac{a_B}{4W}\biggr)\frac{1}{\tau_{RC}}}
{\displaystyle\frac{1}{\tau_g} + 
\biggl(\frac{a_B}{4W}\biggr)\frac{1}{\tau_{RC}} - i\omega}
\biggr\}^{-1}
$$
\begin{equation}\label{A5}
= - \frac{e}{2}\delta V_{\omega} \biggl(\frac{a_B}{4W}\biggr)
\biggl\{\biggl(\frac{a_B}{4W}\biggr)  + \displaystyle
\frac{1}  
{1  - i\omega\tau_r}
\biggr\}^{-1},
\end{equation}
where
 $\tau_{RC} = R_tC_t$ is the time
of the gated section recharging by the tunneling currents,
$R_t$ is the tunneling resistance of the p-n junctions induced by
the negative top gate voltage near the edges of the top gate,
the quantity $\tau_r = \tau_{RC}\tau_g/[\tau_{RC} + (a_B/4W)\tau_g]$
is the characteristic time of the gated section recharging by holes, 
and $C_t = \kappa\,L_t/2\pi\,W$ is the capacitance of the gated
section. 
At $\omega \rightarrow 0$, eq.~(A5) leads to
\begin{equation}\label{A6}
\delta\,\Delta_{\omega}
\simeq - \frac{1}{2}e\delta V_{\omega}
{\cal R}_0, \qquad 
\frac{\partial\,\delta\Delta_{\omega}}{\partial\,
\delta V\omega}\biggr|_{V_d} 
\simeq - \frac{e}{2}{\cal R}_0,
\end{equation}
where 
$$
{\cal R}_0 \simeq 
\frac{(a_B/4W)}{1 + (a_B/4W)} \simeq (a_B/4W),
$$
i.e., coincides with the value given by eq.~(28).
When $\omega \gg \tau_r^{-1}$, eq.~(A5) yields 
\begin{equation}\label{A7}
\delta\,\Delta_{\omega}
\simeq - \frac{1}{2}e\delta V_{\omega}, \qquad 
\frac{\partial\,\delta\Delta_{\omega}}{\partial\,\delta V\omega}\biggr|_{V_d} 
\simeq - \frac{e}{2}.
\end{equation}

\end{document}